\documentstyle[proceedings,numreferences,psfig]{maxent95}

\def\x1{x_{1}}
\def\x{x_{n}}
\def\x1n{x_{1},x_{2},\ldots,x_{n}}
\def\fn{f_{n}}

\newtheorem{theorem}{Theorem}
\newcommand{\node}[1]{
\begin{picture}(10,10)
\put(5,5){\circle{6}}
\put(5,5){\makebox(0,0){#1}}
\end{picture}}
\begin{opening}
\title{CV-NP BAYESIANISM BY MCMC}
\subtitle{Cross Validated Non Parametric Bayesianism by Markov Chain
   Monte Carlo}
\author{Carlos C. Rodriguez}
\institute{Department of Mathematics and Statistics\\
           University at Albany, SUNY\\
           Albany NY 12222, USA\footnote{Email: carlos@math.albany.edu}}
\end{opening}
\runningtitle{CV-NP BY MCMC}

\begin{document}

\title{CV-NP BAYESIANISM BY MCMC}
\author{Carlos C. Rodriguez}

\begin{abstract}
Completely automatic and adaptive non-parametric inference is a pie in the
sky. The frequentist approach, best exemplified by the kernel estimators,
has excellent asymptotic characteristics but it is very sensitive to the
choice of smoothness parameters. On the other hand the Bayesian approach,
best exemplified by the mixture of gaussians models, is optimal given the
observed data but it is very sensitive to the choice of prior. In 1984 the
author proposed to use the Cross-Validated gaussian kernel as the likelihood
for the smoothness scale parameter h, and obtained a closed formula for the
posterior mean of h based on Jeffreys's rule as the prior. The practical
operational characteristics of this bayes' rule for the smoothness parameter
remained unknown for all these years due to the combinatorial complexity of
the formula. It is shown in this paper that a version of the metropolis
algorithm can be used to approximate the value of h producing {\em remarkably%
} good completely automatic and adaptive kernel estimators. A close study of
the form of the cross validated likelihood suggests a modification and a new
approach to Bayesian Non-parametrics in general.


\end{abstract}

\section{Introduction}

A basic problem of statistical inference is to estimate the probability
distribution that generated the observed data. In order to allow the data to 
{\em speak by themselves}, it is desirable to solve this problem with a
minimum of a priori assumptions on the class of possible solutions. For this
reason, the last thirty years have been burgeoning with interest in
nonparametric estimation. The main positive results are almost exclusively
from non-Bayesian quarters, but due to recent advances in Monte Carlo
methods and computer hardware, Bayesian nonparametric estimators are now
becoming more competitive.

This paper aims to extend the idea in {\cite{Rodriguez85}} to a new general
Bayesian approach to nonparametric density estimation. As an illustration
one of the techniques is applied to compute the estimator in {\cite
{Rodriguez85}. In particular it is shown how to approximate the bayes
estimator (for Jeffreys' prior and quadratic loss) for the bandwidth
parameter of a kernel using a version of the metropolis algorithm. The
computer experiments with simulated data clearly show that the bayes
estimator with Jeffreys' prior outperforms the standard estimator produced
by plain Cross-validation. }

\section{Density Estimation by Summing Over Paths}

Any non-Bayesian nonparametric density estimator can be turned into a
Bayesian one by the following three steps:

\begin{enumerate}
\item  Transform the estimator into a likelihood for the smoothness
parameters via the sum over paths technique introduced below.

\item  Put a reference prior on the smoothness parameters.

\item  Use the predictive distribution as the Bayesian estimator obtainable
by Markov Chain Monte Carlo.
\end{enumerate}

The summing over paths method, springs from the simple idea of interchanging
sums and products in the expression for the cross-validated kernel
likelihood (see \cite{Rodriguez85}). Without further reference to its humble
origins, let us postulate the following sequence of functions, 
\begin{equation}
\Phi_{n} = \Phi\left.\left(x_{0},\x1n \right| h \right) \propto \sum_{%
\mbox{{\tiny{all paths}}}} \prod_{j=0}^{n} K_{h}\left(x_{j} -
x_{i_{j}}\right)  \label{eq:phi}
\end{equation}
where $h>0$, $K_{h}(x)=\frac{1}{h}K(x/h)$ with $K$ a density symmetric about 
$0$. We call a vector of indices, $(i_{0},\ldots,i_{n})$ with the property
that $i_{j}\in\{0,\ldots,n\}\setminus\{j\}$, a path (more specifically a
general unrestricted path, see below). The sum above runs over all the $%
n^{n+1}$ possible paths. The functions, $\Phi_{n}$ are defined up to a
proportionality constant independent of the $x_{j}$'s.

Notice that by flipping the sum and the product we get 
\begin{equation}
\frac{1}{n^{n+1}} \sum_{\mbox{{\tiny{all paths}}}}\prod_{j=0}^{n}
K_{h}\left(x_{j} - x_{i_{j}}\right) = f_{-0,n}(x_{0}) f_{-1,n}(x_{1}) \cdots
f_{-n,n}(x_{n})  \label{eq:prod}
\end{equation}
where, 
\begin{equation}
f_{-j,n}(x_{j}) = \frac{1}{n} \sum_{\stackrel{\scriptstyle i_{j} = 0}{i_{j}
\neq j}}^{n} K_{h}\left(x_{j} - x_{i_{j}}\right).  \label{eq:ker}
\end{equation}
Thus, $f_{-j,n}(x_{j})$ is nothing but the kernel density estimator of $%
f(x_{j})$ based on all the data except the jth observation $x_{j}$. Under
mild regularity conditions the kernel estimator is known to converge (in
every conceivable way) provided that $h = h_{n}$ is taken as a function of $n
$ such that, $h_{n}\rightarrow 0$ and $n h_{n}\rightarrow \infty$ as $%
n\rightarrow\infty$.

The $\Phi_{n}$'s can be used as a {\em universal} method for attaching a
class of exchangeable one parameter models to any set of observations. The
positive scalar parameter $h$ is the only free parameter, and different
models are obtained by changing the kernel function $K$.

These empirical parametric models are invariant under relabeling of the
observations (i.e. they are exchangeable) but they do not model the
observations as independent variables. Rather, these models introduce a
pattern of correlations for which there is a priori no justification. This
suggest that there might be improvements in performance if the sum is
restricted to special subsets of the set of all $n^{n+1}$ paths. Three of
these modifications are mentioned in the following section.

Notice also that the ability of the $\Phi_{n}$ to adapt comes at the expense
of regularity. These models are always non-regular. If the kernel has
unbounded support then $\Phi_{n}$ integrates to infinity but the conditional
distribution of $x_{0}$ given $x_{1},\ldots,x_{n}$ and $h$ is proper. When
the kernel has compact support the $\Phi_{n}$ are proper but still
non-regular since their support now depends on $h$.

The above recipe would have been a capricious combination of symbols if it
not were for the fact that, under mild regularity conditions, these models 
{\em adapt} to the form of the true likelihood as $n$ increases.

As a function of $x_{0}=x$, the $\Phi_{n}$ have the following asymptotic
property,

\begin{theorem}
If $\x1n$ are iid observations from an unknown pdf $f$ which is
continuous a.s. and $h=h_{n}$ is taken as a function of $n$ such that, $%
h_{n}\rightarrow 0$ and $nh_{n}\rightarrow \infty $ as $n\rightarrow \infty $%
, then, 
\begin{equation}
\frac{\Phi _{n}}{\int \Phi _{n}dx_{0}}=f_{-0,n}(x)+o(1)=f(x)+o(1).
\label{eq:th1}
\end{equation}
where the little $o$ is taken in probability as $n\rightarrow \infty $
\end{theorem}

{\bf Proof} (sketch only)\newline
Just flip the sum and the product to get again,

\begin{equation}
\frac{1}{n^{n+1}} \sum_{\mbox{{\tiny{all paths}}}}\prod_{j=0}^{n}
K_{h}\left(x_{j} - x_{i_{j}}\right) = f_{-0,n}(x) f_{-1,n}(x_{1}) \cdots
f_{-n,n}(x_{n})
\end{equation}
Under the simple regularity conditions of the theorem, the kernel estimator
is known to converge in probability as $n\rightarrow\infty$. However, even
though $x_{0}$ appears in all of the $n+1$ factors, and their number goes to
infinity, still all the factors are converging to the value of the true
density at the given point. Therefore the theorem follows.$\Box$

It is worth noticing that the above theorem is only one of a large number of
results that are readily available from the thirty years of literature on
density estimation. In fact under appropriate regularity conditions the
convergence can be strengthen to be pointwise a.s., uniformly a.s., or
globally a.s. in $L_{1}$, or $L_{2}$.

\section{Paths, Graphs and Loops}

\begin{figure}[tbp]
\begin{center}
\begin{picture}(40,40)
\put(10,10){\node{$x_{0}$}}
\put(18,18){\vector(1,1){5}}
\put(20,20){\node{$x_{2}$}}
\put(30,25){\vector(1,0){4}}
\put(32,20){\node{$x_{1}$}}
\put(38,22){\vector(-1,-1){4}}
\put(25,10){\node{$x_{3}$}}
\put(31,18){\vector(-1,1){4}}
\end{picture}
\end{center}
\caption{The graph of $(2,3,1,2)$}
\label{fig:d0.1}
\end{figure}
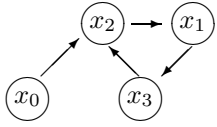

Each of the $n^{n+1}$ paths $(i_{0},\ldots,i_{n})$ can be represented by a
graph with nodes $x_{0},\ldots,x_{n}$ and edges from $x_{j}$ to $x_{k}$ if
and only if $i_{j} = k$. Here are some graphs for paths with $n=3$. For
example, the path $(2,3,1,2)$ is given a probability proportional to 
\begin{equation}
K_{h}(x_{0}-x_{2})K_{h}(x_{1}-x_{3})K_{h}(x_{2}-x_{1})K_{h}(x_{3}-x_{2})
\end{equation}
and represented by the graph in figure [\ref{fig:d0.1}]. Let's call it a
1-3-loop. 
The path $(1,2,3,0)$ is the single ordered loop of size four (a 4-loop), $%
(3,0,1,2)$ is the same loop backwards (also a 4-loop), $(2,3,0,1)$ are two
disconnected loops (a 2-2-loop) and $(1,0,0,0)$ is connected and contains a
loop of size two with $x_{0}$ and $x_{1}$ in it (a 1-1-2-loop). Draw the
pictures!

The classification of paths in terms of number and size of loops appears
naturally when trying to understand how $\Phi_{n}$ distributes probability
mass among the different paths. To be able to provide simple explicit
formulas let us take $K$ in the definition of $\Phi_{n}$ to be the standard
gaussian density, i.e. from now on we take, 
\begin{equation}
K(x) = \frac{1}{\sqrt{2\pi}}\exp{\left(\frac{-x^{2}}{2}\right)}.
\end{equation}
The gaussian kernel has unbounded support and that makes the total integral
of each path to diverge. Thus, the partition function 
\begin{eqnarray}
Z & = & \int\Phi_{n}dx_{0}\ldots dx_{n} \\
& = & \sum_{\mbox{{\tiny{all paths}}}}\int\prod_{j=0}^{n}K_{h}\left(x_{j} -
x_{i_{j}}\right) dx_{0}\ldots dx_{n}  \label{eq:Z}
\end{eqnarray}
is the sum of infinities and it also diverges. Recall that this anomaly is
the price we need to pay for using a model with a finite number of free
parameters (only one in this case) and hoping to still adapt to the form of
the true likelihood as $n\rightarrow\infty$. Even though the value of $Z$ is
in fact $\infty$ we can still write (formally) a useful decomposition that
will help explain how the $\Phi_{n}$'s adapt and how to modify the set of
paths to improve the convergence. We first need the following simple
property of gaussians, 
\begin{equation}
\int K_{a}(x-y) K_{b}(y-z) dy = K_{\sqrt{a^{2}+b^{2}}}(x-z).  \label{eq:kakb}
\end{equation}
This can be shown by straight forward integration after completing the
square. Now notice that whatever the value of the integrals appearing in
equation (\ref{eq:Z}) that value only depends on the type of loop that is
being integrated. For this reason we omit the integrand and simply denote
the value of the integral with the integral sign and the type of loop. With
this notation we have,

\begin{theorem}
\begin{equation}
\int_{m_{1}-m_{2}-\ldots -m_{k}-\makebox{loop}}=\left( \int_{m_{1}-%
\makebox{loop}}\right) \left( \int_{m_{2}-\makebox{loop}}\right) \ldots
\left( \int_{m_{k}-\makebox{loop}}\right)   \label{eq:m1m2}
\end{equation}
More over, $\int_{1-\makebox{loop}}=1$ and for $m>1$, 
\begin{equation}
\int_{m-\makebox{loop}}=\frac{1}{\sqrt{2\pi }}\frac{h^{-1}}{\sqrt{m}}L
\label{eq:mloop}
\end{equation}
where we write formally $L=\int dx$.
\end{theorem}

{\bf Proof} \newline
Equation (\ref{eq:m1m2}) follows from Fubini's theorem. To get (\ref
{eq:mloop}) use Fubini's theorem and apply (\ref{eq:kakb}) each time to
obtain, 
\begin{eqnarray*}
\int_{m-\makebox{loop}} & = & \int K_{h}(x_{0}-x_{1})K_{h}(x_{1} - x_{2})
K_{h}(x_{2}-x_{3}) \ldots K_{h}(x_{m-1}-x_{0}) dx_{0}dx_{1}\ldots dx_{m-1} \\
& = & \int K_{\sqrt{2}h}(x_{0}-x_{2}) \ldots K_{h}(x_{m-1}-x_{0})
dx_{0}dx_{2}\ldots dx_{m-1} \\
& \ldots & \\
& = & \int K_{\sqrt{m-1}h}(x_{0}-x_{m-1})K_{h}(x_{m-1}-x_{0}) dx_{0}dx_{m-1}
\\
& = & \int K_{\sqrt{m}h}(0) dx_{0} = \frac{1}{\sqrt{2\pi}} \frac{h^{-1}}{%
\sqrt{m}} L
\end{eqnarray*}
$\Box$\newline
Hence, by splitting the sum over all paths into, 
\[
\sum_{\mbox{{\tiny{all paths}}}} = \sum_{\mbox{{\tiny{2-2...-2-loops}}}} + 
\sum_{\mbox{{\tiny{1-3-2...-2-loops}}}} + \ldots +  \sum_{%
\mbox{{\tiny{(n+1)-loops}}}} 
\]
and applying the previous theorem we obtain, 
\begin{equation}
Z = N_{2-2\ldots{-2}} \left(\frac{1}{\sqrt{2\pi}} \frac{h^{-1}}{\sqrt{2}}
L\right)^{(n+1)/2} + \ldots + N_{n+1}\left(\frac{1}{\sqrt{2\pi}} \frac{h^{-1}%
}{\sqrt{n+1}} L\right)^{1}  \label{eq:zz}
\end{equation}
where for simplicity we have assumed that $n$ is odd and we denote by $%
N_{m_{1}-\ldots -m_{k}}$ the total number of $m_{1}-\ldots -m_{k}-%
\mbox{loops}$. Using simple combinatorial arguments it is possible to write
explicit formulas for the number of loops of each kind. The important
conclusion from the decomposition (\ref{eq:zz}) is that even though the $%
\Phi_{n}$ appear to be adding equally over all paths, in reality they end up
allocating almost all the probability mass on paths with maximally
disconnected graphs. This is not surprising. This is the reason why there is
consistency when assuming iid observations. There is a built in bias towards
independence. The bias can be imposed explicitly on the $\Phi_{n}$ by
restricting the paths to be considered in the sum. Here are three examples:

\begin{description}
\item[loops:]  \ \newline
Only paths $(i_{0},\ldots ,i_{n})$ that form a permutation of the integers $%
\{0,1,\ldots ,n\}$ are considered.

\item[$2-2-\ldots-2-$loops:]  \ \newline
Only maximally disconnected paths are considered.

\item[QM:]  \ \newline
Paths as above but use $\left| \Phi _{n}\right| ^{2}$ instead of $\Phi _{n}$
as the joint likelihoods.
\end{description}

Preliminary simulation experiments seem to indicate that only maximally
disconnected paths are not enough and that all the loops are too many. The
QM method has all the maximally disconnected paths but not all the loops
(e.g. with $n=5$ the 3-3-loops can not be reached by squaring the sum of
2-2-2-loops) so it looks like the most promising among the three. What is
more interesting about the QM method is the possibility of using kernels
that can go negative or even complex valued. More research is needed since
very little is known about the performance of these estimators.

\section{Estimation by MCMC}

We show in this section how to approximate the predictive distribution and
the bayes rule for the smoothness parameter by using Markov Chain Monte
Carlo techniques.

\subsection{Posterior mean of the bandwidth}

Apply bayes' theorem to (\ref{eq:phi}) to obtain the posterior distribution
of $h$, 
\begin{equation}
\pi(h|x,\x1n) = \frac{\Phi(x,\x1n|h) \pi(h)} {\int_{0}^{\infty}%
\Phi(x,\x1n|\tau) \pi(\tau) d\tau}  \label{eq:post}
\end{equation}
where $\pi$ is a function of $h$. It is worth noticing that $\pi$ is {\bf not%
} the prior on $h$. It is only the part of the prior on $h$ that we can
manipulate. Recall that $\Phi_{n}$ integrates to the function of $h$ given
by (\ref{eq:zz}) so effectively the prior that is producing (\ref{eq:post})
is, 
\begin{equation}
\Pi(h) \propto \frac{\pi(h)}{h^{(n+1)/2}}  \label{eq:prior}
\end{equation}
The posterior mean is then given by, 
\begin{equation}
E(h|x,\x1n) = \hat{h}_{x} = \frac{\int_{0}^{\infty} h \Phi(x,\x1n|h)
\pi(h) dh} {\int_{0}^{\infty}\Phi(x,\x1n|h) \pi(h) dh}  \label{eq:pmean}
\end{equation}
Equation (\ref{eq:pmean}) provides a different estimator for each value of $x
$. To obtain a single global estimate for $h$ just erase the $x$'s from (\ref
{eq:pmean}) and change $n$ to $n-1$ in the formulas below. When $K$ is the
univariate gaussian kernel and $\pi(h) = h^{-\delta}$ equation (\ref
{eq:pmean}) simplifies to: 
\begin{equation}
\hat{h}_{x} = C_{n,\delta} \frac{\sum_{\mbox{{\tiny{all paths}}}}\alpha(%
\underline{i})s(\underline{i})} {\sum_{\mbox{{\tiny{all paths}}}}\alpha(%
\underline{i})}  \label{eq:hhat}
\end{equation}
where, 
\begin{equation}
C_{n,\delta} = \frac{1}{\sqrt{2}} \frac{\Gamma\left(\frac{n+\delta-1}{2}%
\right)} {\Gamma\left(\frac{n+\delta}{2}\right)}  \label{eq:cnd}
\end{equation}
$\underline{i}=(i_{0},\ldots,i_{n})$ is a path, $\alpha = s^{-(n+\delta)}$
and 
\begin{equation}
s^{2}(\underline{i}) = \sum_{j=0}^{n}(x_{j}-x_{i_{j}})^{2}.
\label{eq:sigma2}
\end{equation}
Equation (\ref{eq:hhat}) follows from two applications of the formula, 
\begin{equation}
\int_{0}^{\infty}h^{-(\beta+1)}\exp{\left\{-\frac{s^{2}}{2h^{2}}\right\}} dh
= \frac{1}{2}2^{\beta/2}\Gamma(\beta/2) s^{-\beta}  \label{eq:intb}
\end{equation}

\subsubsection{Bandwidth by Metropolis}

To approximate equation (\ref{eq:hhat}) we use the fact that the ratio of
the two sums is the expected value of a random variable that takes the value 
$s(\underline{i})$ on the path $\underline{i}$ which is generated with
probability proportional to $\alpha(\underline{i})$. The following version
of the Metropolis algorithm produces a sequence of averages that converge to
the expected value, \newline
\newline
{\bf Algorithm}

\begin{verse}
0) Start from somewhere \newline
$\underline{i} \leftarrow (1,2,\ldots,n,0)$ \newline
$s2 \leftarrow \sum_{j=0}^{n} (x_{j} - x_{i_{j}})^{2}$ \newline
$\alpha \leftarrow (s2)^{-(n+\delta)/2}$ \newline
$N \leftarrow 0$, sum $\leftarrow 0$, ave $\leftarrow 0$  \newline

1) Sweep along the components of $\underline{i}$ \newline
for k from 0 to n do \newline

\{ \newline
${i}^{\prime}_{k} \leftarrow$ Uniform on $\{0,\ldots,n\}\setminus\{k,i_{k}\}$
\newline
$\Delta_{k} \leftarrow (x_{{i}^{\prime}_{k}}-x_{i_{k}})  (x_{{i}%
^{\prime}_{k}}+x_{i_{k}}-2x_{k})$ \newline
$s2^{\prime} \leftarrow s2 + \Delta_{k}$ \newline
$\alpha^{\prime} \leftarrow (s2^{\prime})^{-(n+\delta)/2}$ \newline
$R \leftarrow \alpha^{\prime}/\alpha$ \newline
if $R > 1$ or Unif[0,1] $< R$ then $\{i_{k} \leftarrow {i}^{\prime}_{k},
s2\leftarrow s2^{\prime},  \alpha \leftarrow \alpha^{\prime}\}$ \newline
\} \newline

2) Update the estimate for the average, \newline
sum $\leftarrow$ sum $+\sqrt{s2}$ \newline
$N\leftarrow N+1$ \newline
ave $\leftarrow \mbox{sum}/N$ \newline
goto 1)\newline
\end{verse}

\subsection{The Predictive Distribution by Gibbs}

To sample from the predictive distribution, $f(x|\x1n)$ we use Gibbs to
sample from the joint distribution, $f(x,h|\x1n)$. Hence, we only need to
know how to sample from the two conditionals, a) $f(x|h,\x1n)$ and b) $%
\pi(h|x,\x1n)$. To sample from a) we use the fact that this is (almost)
the classical kernel so all we need is to generate from an equiprobable
mixture of gaussians. To get samples from b) just replace the gaussian
kernel into the numerator of equation (\ref{eq:post}) to obtain, for $%
\pi(h)\propto h^{-\delta}$, 
\begin{equation}
\pi(h|x,\x1n) \propto \sum_{\mbox{{\tiny{all paths}}}} h^{-(n+\delta+1)}
\exp{\left\{-\frac{s^{2}}{2h^{2}}\right\}}.  \label{eq:post1}
\end{equation}
The integral with respect to $h$ of each of the terms being added in (\ref
{eq:post1}) is proportional to $s^{-(n+\delta)}$ (see (\ref{eq:intb})).
Thus, by multiplying and dividing by this integral each term, we can write, 
\begin{equation}
\pi(h|x,\x1n) \propto \sum_{\mbox{{\tiny{all paths}}}} \alpha(\underline{i%
}) \pi_{s(\underline{i})}(h)  \label{eq:post2}
\end{equation}
where $\alpha(\underline{i}) = (s(\underline{i}))^{-(n+\delta)}$ as before
and, 
\begin{equation}
\pi_{s}(h) \propto h^{-(n+\delta+1)} \exp{\left\{-\frac{s^{2}}{2h^{2}}%
\right\}}  \label{eq:pish}
\end{equation}
From the change of variables theorem it follows that if $y$ is Gamma$(\frac{%
n+\delta}{2},1)$ then $h = \sqrt{\frac{s^{2}}{2y}}$ follows the distribution
(\ref{eq:pish}). This shows that the posterior distribution of $h$ is a
mixture of transformed gamma distributions. This mixture can be generated by
a simple modification to the algorithm used to get the posterior mean of $h$.

\section{Experiments on simulated and real data}

\begin{figure}[tbp]
\begin{center}
\psfig{figure=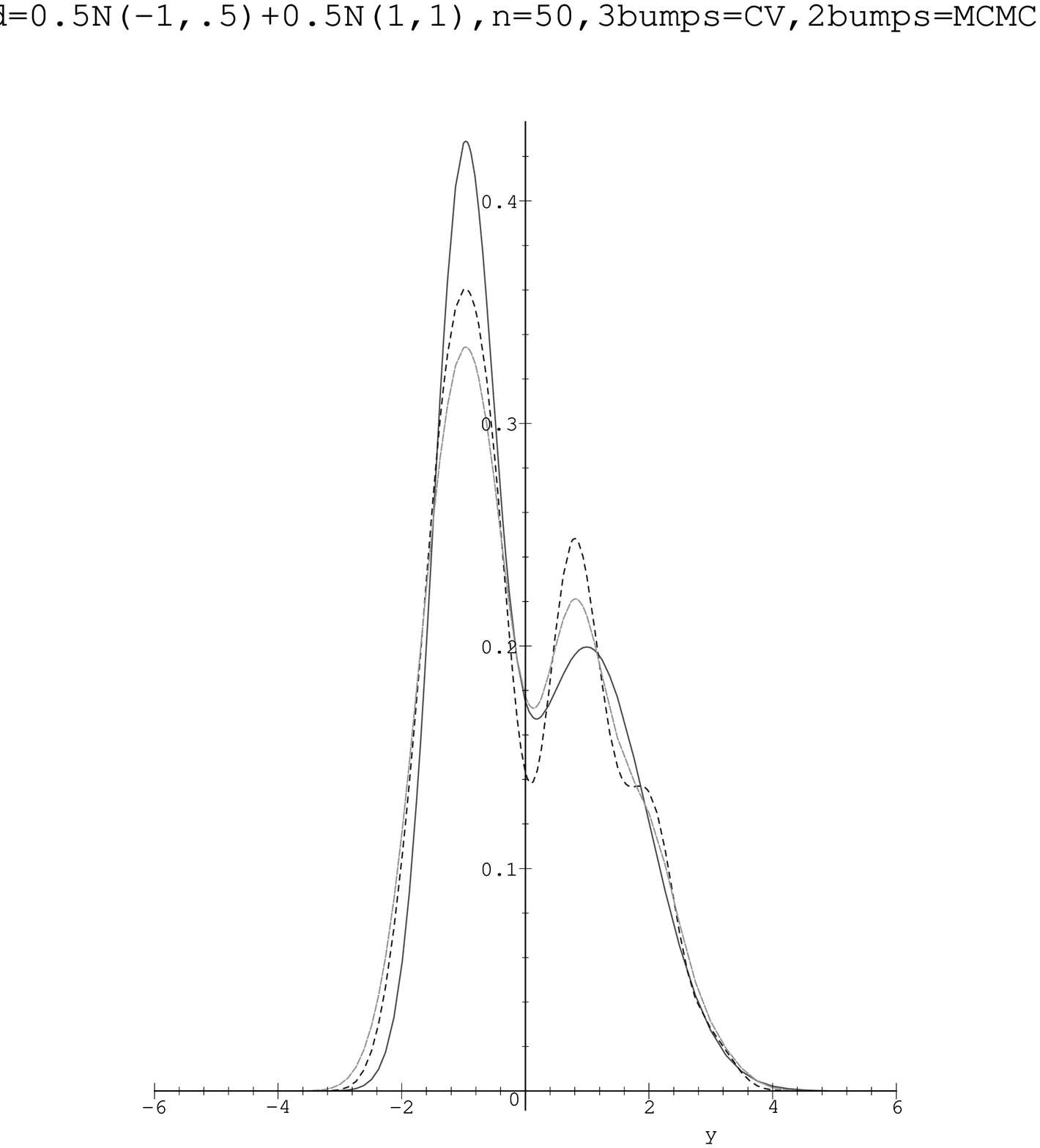,width=3.5in}
\end{center}
\caption{Posterior mean of global h vs plain cross-validation}
\label{fig:d0.3}
\end{figure}

I have coded (essentially) the algorithm described in the previous section
in MAPLE and tested it dozens of times on simulated data for computing a
global value for the smoothness parameter $h$. All the experiments were
carried out with $\delta = 1$ i.e. with $\pi(h)=h^{-1}$ on mixtures of
gaussians. The experiments clearly indicate that the global value of $h$
provided by the MCMC algorithm produce a kernel estimator that is either
identical to plain likelihood cross-validation or clearly superior to it
depending on the experiment. A typical run is presented in figure [\ref
{fig:d0.3}] where the true density and the two estimators from 50 iid
observations are shown. The MAPLE package used in the experiments is
available at {\em http://omega.albany.edu:8008/npde.mpl}.

\begin{figure}[tbp]
\begin{center}
\psfig{figure=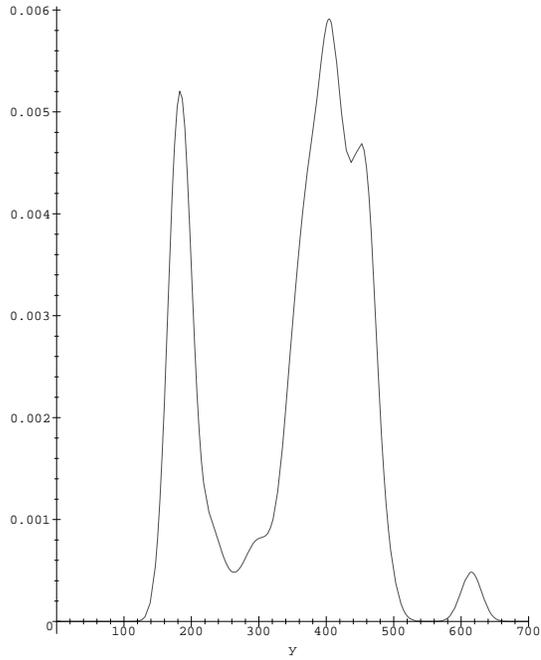,width=3.5in}
\end{center}
\caption{Estimate for the Old Faithful geyser data, h = 14.217}
\label{fig:d0.2}
\end{figure}

For comparison with other density estimators in the literature we show in
figure [\ref{fig:d0.2}] the estimate for the complete set of 109
observations of the Old Faithful geyser data. These data are the 107
observations in {\cite{Silverman86}} plus the two outliers 610 and 620. This
is a standard gaussian kernel with the global value of $h=14.217$ chosen by
the MCMC algorithm. 

\section{Conclusions}

There is nothing special about dimension one. 
Only minor cosmetic changes (of the kind: replace $h$ to $h^{d}$ 
in some formulas) are needed to include the multivariate case, i.e.
the case when the $x_{j}$'s are d-dimensional vectors instead 
of real variables.

Very little is known about these estimators 
beyond of what it is presented in this paper, 
In particular nothing is known about rates of convergence. 
There are many avenues to explore with theory and with simulations
but clearly the most interesting and promissing
open questions are those related to the performance of the QM method above.

\bibliographystyle{maxent95}
\bibliography{CV-NP}

\begin{thebibliography}{1}

\bibitem{Rodriguez85}
C.~C. Rodriguez, ``On the estimation of the bandwidth parameter using a non
  informative prior,'' in {\em Proceedings of the 45th session of the ISI},
  No.~1, pp.~207--208, International Statistical Institute, August 1985.

\bibitem{Silverman86}
B.~W. Silverman, {\em Density Estimation: for statistics and data analysis},
  Monographs on Statistis and Applied Probability, Chapman and Hall, 1986.

\end{thebibliography}

\end{document}